\documentclass[alpha-refs]{wiley-article}

\usepackage{graphicx}
\usepackage[space]{grffile}
\usepackage{latexsym}
\usepackage{textcomp}
\usepackage{longtable}
\usepackage{tabulary}
\usepackage{booktabs,array,multirow}
\usepackage{amsfonts,amsmath,amssymb}
\usepackage{natbib}
\usepackage{url}
\usepackage{hyperref}
\hypersetup{colorlinks=false,pdfborder={0 0 0}}
\usepackage{etoolbox}
\usepackage{placeins}
\newif\iflatexml\latexmlfalse

\AtBeginDocument{\DeclareGraphicsExtensions{.pdf,.PDF,.eps,.EPS,.png,.PNG,.tif,.TIF,.jpg,.JPG,.jpeg,.JPEG}}

\usepackage[utf8]{inputenc}
\usepackage[english]{babel}

\usepackage{siunitx}
\usepackage[nolists,nomarkers]{endfloat}

\usepackage{setspace}
\DeclareMathOperator*{\argmax}{arg\,max}

\iflatexml

\else

\newcommand{\xs}[1]{\textbf{x}_{#1}}

\paperfield{Geophysical Prospecting}
\corraddress{Johno van IJsseldijk, Department of Geoscience \& Engineering, Delft University of Technology, 2628 CN, Delft, The Netherlands}
\corremail{J.E.vanIJsseldijk@tudelft.nl}
\fundinginfo{European Research Council (ERC) under the European Union's Horizon 2020 research and innovation programme (Grant Agreement 742703)}
\fi

\papertype{Original Article}

\title{Time-lapse applications of the Marchenko method on the Troll field}

\author[1]{Johno van IJsseldijk}
\author[2]{Joeri Brackenhoff}
\author[1]{Jan Thorbecke}
\author[1]{Kees Wapenaar}

\affil[1]{Delft University of Technology, Department of Geoscience and Engineering}
\affil[2]{Quantairra Research and Development Services B.V.}

\makeatletter
\let\oldtheequation\theequation
\renewcommand\tagform@[1]{\maketag@@@{\ignorespaces#1\unskip\@@italiccorr}}
\renewcommand\theequation{(\oldtheequation)}
\makeatother

\runningauthor{Marchenko-based monitoring on field data \hspace{.2cm} |\hspace{.2cm} van IJsseldijk et al.}

\begin{document}

\sloppy
\begin{singlespace}
\maketitle
\end{singlespace}

\selectlanguage{english}
\begin{abstract}
\begin{singlespace} \vspace*{-\baselineskip}
\normalfont\large
The data-driven Marchenko method is able to redatum wavefields to arbitrary locations in the subsurface, and can, therefore, be used to isolate zones of specific interest. This creates a new reflection response of the target zone without interference from over- or underburden reflectors. Consequently, the method is well suited to obtain a clear response of a subsurface reservoir, which can be advantageous in time-lapse studies. The isolated responses of a baseline and monitor survey can be more effectively compared, hence  the retrieval of time-lapse characteristics is improved. This research aims to apply Marchenko-based isolation to a time-lapse marine dataset of the Troll field in Norway in order to acquire an unobstructed image of the primary reflections, and retrieve small time-lapse traveltime difference in the reservoir. It is found that the method not only isolates the primary reflections but can also estimate internal multiples outside the recording time. Both the primaries and the multiples can then be utilised to find time-lapse traveltime differences. More accurate ways of time-lapse monitoring will allow for a better understanding of dynamic processes in the subsurface, such as  observing saturation and pressure changes in a reservoir or monitoring underground storage of hydrogen and CO$_2$.  
\end{singlespace}
\end{abstract}%

\large
\section*{Introduction}
Time-lapse seismic has become increasingly important to monitor fluid flows and geomechanical changes in subsurface reservoirs, such as observing pressure and saturation changes \citep{Johnston1998,Landro2001,Dadashpour2007}, monitoring CO$_2$ storage sites \citep{Chadwick2010,Pevzner2011,Ivandic2018} and assessing compaction and subsidence \citep{Barkved2005,Hatchell2005}. Typically, these studies compare an initial baseline study followed by one or more monitor studies. From these studies small differences in amplitude \citep{Landro2001}, in traveltime \citep[][and this work]{Landro2004,Macbeth2019} or in a combination of both \citep{Tura2005,Trani2011} can be observed. These changes can be retrieved by independently creating an image for both the baseline and monitor study, and subtracting these images from one another to find the time-lapse differences. This subtraction highlights the dynamic differences, for example the fluid flow in the reservoir, while removing the static part from the data, such as the time invariant geology \citep{Lumley2001}. \\
A common technique to retrieve time-lapse time differences is by cross-correlating the signal of the baseline and monitor surveys. This can either be done on picked events shared by both surveys or to the complete data sets all at once \citep{Macbeth2020}. \citet{Snieder2002} show that due to multiple scattering, correlations of the coda of a signal can display larger time-lapse effects compared to correlating first arrivals. This technique, called coda wave interferometry, can be applied on laboratory scale to core samples \citep{Singh2019} as well as on field scales to monitor temporal changes in a volcano \citep{Gret2005}. \citet{WapenaarIJsseldijk2020} introduce a novel methodology to clearly identify the reservoir response from a seismic survey using Marchenko-based isolation as well as to improve the detectability of the traveltime changes by correlating reservoir-related internal multiples, akin to the principle of coda wave interferometry. \\
At the base of this new methodology are the Marchenko equations, which allows for a data-driven  redatuming of the seismic wavefield to an arbitrary focal point in the subsurface \citep{Slob2014,wapenaar2014}. Since all orders of internal multiples are accounted for, the redatumed wavefields are free from any interactions of the overburden, hence providing an unobstructed view of the primary reflections of the reservoir when a focal level just above the reservoir is chosen. The Marchenko equations can then be applied a second time to the newly found reflection response to also remove underburden interactions \citep{WapenaarAndStaring2018}. If the second focal depth is chosen just below the reservoir, the final result has effectively isolated all primaries and multiples of the reservoir. This isolated response can then be used to more accurately retrieve time-lapse traveltime shifts due to changes in the reservoir by cross-correlating baseline and monitor responses \citep{vanIJsseldijk2021,vanIJsseldijk2023}. \\
Here, the aim is to apply the Marchenko method to marine time-lapse datasets of the Troll Field and retrieve accurate time-lapse traveltime shifts. In order to do this we first review the theory of isolating the reservoir response and how to extract time-lapse traveltime differences from the primary and multiple reflections. Next, the Troll Field data are introduced; before the methodology can be applied a number of preprocessing steps and limitations of the data need to be considered. After properly preparing both the baseline and monitor surveys, Marchenko-based isolation is used to enhance the reservoir response  for time-lapse analysis. Finally, the traveltime differences related to the reservoir are calculated from suitable primaries and multiples. 

\par\null

\section*{Theory}
This section briefly reviews the theory of Marchenko-based isolation of the reservoir response from the full reflection response. After applying this isolation to both a baseline and monitor study, the traveltime differences inside the reservoir can be more accurately calculated, as described in the second part of this section. A full derivation of the Marchenko method is beyond the scope of this paper; instead only relevant equations are discussed here. \citet{Wapenaar2021} provide a more thorough derivation and background on the Marchenko method. 

\subsection*{Marchenko-based isolation}
\begin{figure}[tb!]
\includegraphics[width=\textwidth]{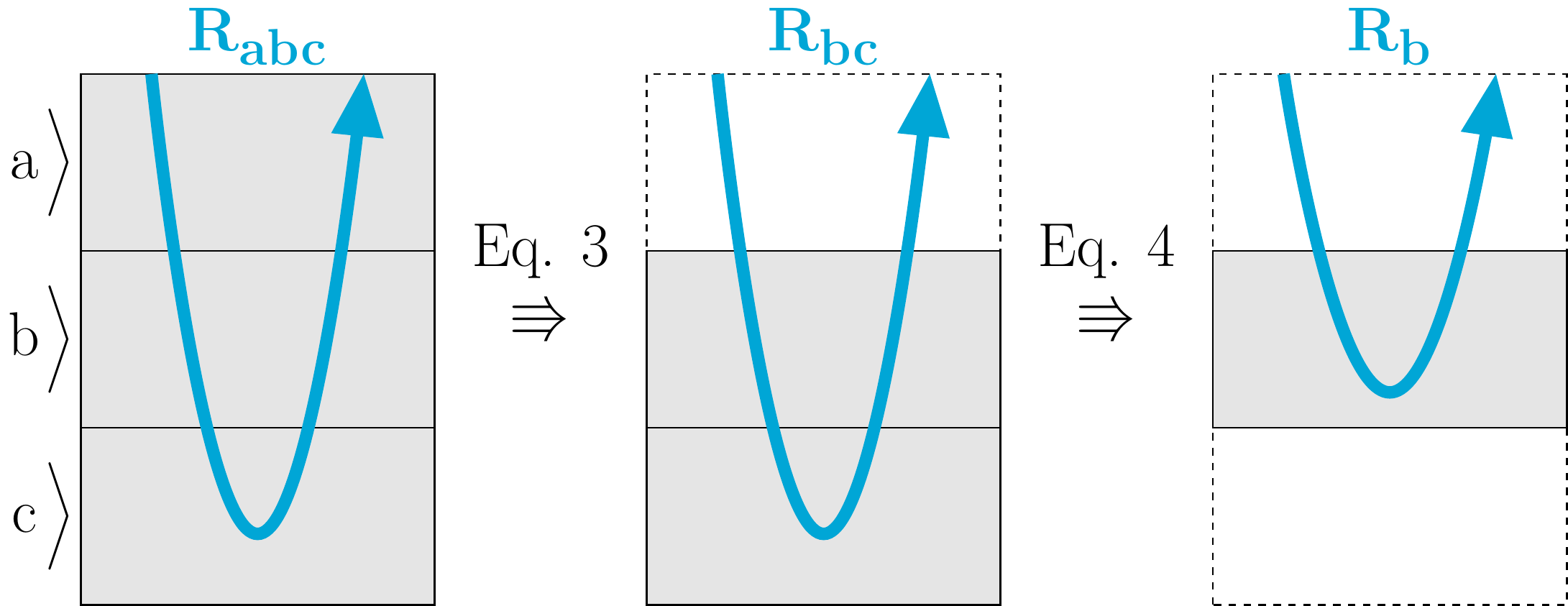}
\caption{Graphic showing the concept of Marchenko-based isolation. The medium is divided in 3 units; overburden $a$, target zone $b$ and underburden $c$. First, the overburden is removed from the response as described in equation 3. Next, the underburden is removed, leaving only the target zone with the reservoir response. }
\label{fig:concept}
\end{figure}

The Marchenko method relies on two Green's function representations that relate the extrapolated Green's functions ($U^{-,\pm}$) to the extrapolated focusing functions ($v^\pm$) via the reflection response $R(\xs{R},\xs{S},t)$. In this notation the first and second coordinate describe the receiver and source position, respectively, and $t$ denotes time. The superscripts ${-,\pm}$ represent an up-going receiver field from a up ($-$) or down-going ($+$) source field. The focusing functions are defined in a truncated medium, which is the same as the actual medium above an arbitrary focal level and homogeneous below this level. In the actual medium the focusing functions let the wavefield converge to the focal point, creating a virtual source that produces the Green's functions between the focal depth and the surface. Both the focusing and Green's functions are extrapolated from the focal depth to the surface, so that the coordinates of all the functions are located at acquisition surface $\mathbb{S}_0$. The focusing of the wavefield in the actual medium is then described by the following equations \citep{vanderNeut2016}:
\begin{equation}
\label{eqn:mar1}
U^{-,+}(\xs{R},\xs{S}',t) + v^-(\xs{R},\xs{S}',t) =  \int_{\mathbb{S}_0} R(\xs{R},\xs{S},t) * v^+(\xs{S},\xs{S}',t)d\xs{S},
\end{equation}
and
\begin{equation}
 \label{eqn:mar2}
U^{-,-}(\xs{R},\xs{S}',-t) + v^+(\xs{R},\xs{S}',t) =  \int_{\mathbb{S}_0} R(\xs{R},\xs{S},-t) * v^-(\xs{S},\xs{S}',t)d\xs{S}.
\end{equation}
Here, $*$ denotes a convolution and the right-hand side integrates over the source positions $\xs{S}$ at the acquisition surface $\mathbb{S}_0$. These two equations have four unknowns, hence, to solve the equations, an additional causality constraint is introduced, which takes advantage of the fact that the focusing and Green's functions are separable in time \citep{wapenaar2014}. In order to achieve this separation an estimate of the two-way traveltime (twt) from $\mathbb{S}_0$ to the focal depth and back is required. This estimate can for example be obtained from a smooth velocity model. By limiting Equations \ref{eqn:mar1} and \ref{eqn:mar2} between $t=0$ s and this twt, the Green's functions in the left-hand side vanish \citep{vanderNeut2016}; the resulting equations are known as the extrapolated Marchenko equations, which now only contain two unknowns and can, therefore, be solved iteratively \citep{Thorbecke2017} or by inversion \citep{vanderNeut2015}. \\
Next, the subsurface is divided in three units; overburden $a$, target zone $b$ and underburden $c$ as shown in Figure \ref{fig:concept}. The over- and underburden contain undesirable responses, whereas the target zone contains the reservoir of interest for the time-lapse study. The primary and multiple reflections of the over- and underburden can be removed using a twofold Marchenko-based strategy, leaving a reflection response only containing events from the target zone. First, Equations \ref{eqn:mar1} and \ref{eqn:mar2} are used to find the extrapolated Green's functions with a focal level between overburden $a$ and target zone $b$.  Using these Green's functions, a reflection response free of overburden interactions can be acquired by solving \citep{Wapenaar2021}:
\begin{equation}
\label{eqn:redatum}
U^{-,+}_{a|bc}(\xs{R},\xs{S}',t) =  -\int_{\mathbb{S}_0} U^{-,-}_{a|bc}(\xs{R},\xs{R}',t) * R_{bc}(\xs{R}',\xs{S}',t) d \xs{R}'.
\end{equation}
The subscript $a|bc$ denotes that the extrapolated Green's functions are retrieved from the full reflection response ($R_{abc}$) with a focal depth between units $a$ and $b$. The reflection response $R_{bc}$ is retrieved by a multi-dimensional deconvolution \citep[MDD,][]{Broggini2014}, and contains all primary and multiple reflections from $b$ and $c$, but none from overburden $a$ (Figure \ref{fig:concept}b). This new reflection response can then be used to find focusing functions below the target zone, which in turn are used to find the reflection response that only contains target zone events \citep{WapenaarAndStaring2018}:
\begin{equation}
\label{eqn:isolation}
v^-_{b|c}(\xs{R},\xs{S}',t) = 
 \int_{\mathbb{S}_0} v^+_{b|c} (\xs{R},\xs{R}',t) * R_{b}(\xs{R}',\xs{S}',t) d \xs{R}'.
\end{equation}
The subscript $b|c$ denotes that the extrapolated focusing functions are retrieved from the reflection response without overburden ($R_{bc}$) using a focal depth at the interface of units $b$ and $c$. Equation \ref{eqn:isolation} directly follows from the definition of a focusing function in the truncated medium \citep{Wapenaar2021}. Once again, the isolated reflection response $R_b$ can be retrieved from this equation by MDD, and consists solely of the reflections (primary and multiple) from the target zone. \citet{vanIJsseldijk2023} demonstrate how the internal multiples in the target area $b$ can be artificially enhanced by increasing the amplitude of the coda of the downgoing focusing function $v^+$, in order to benefit the identification of the multiples and the extraction of time-lapse traveltime differences. Figure \ref{fig:concept} summarises the twofold approach to isolate the reservoir response from a full reflection response. This process is applied to the baseline as well as the monitor study, to acquire the isolated reservoir responses in both studies. These isolated responses are then used to find the time-lapse traveltime differences inside the reservoir. 

\subsection*{Extracting travel-time differences}
\begin{figure}[tb!]
\includegraphics[width=\textwidth]{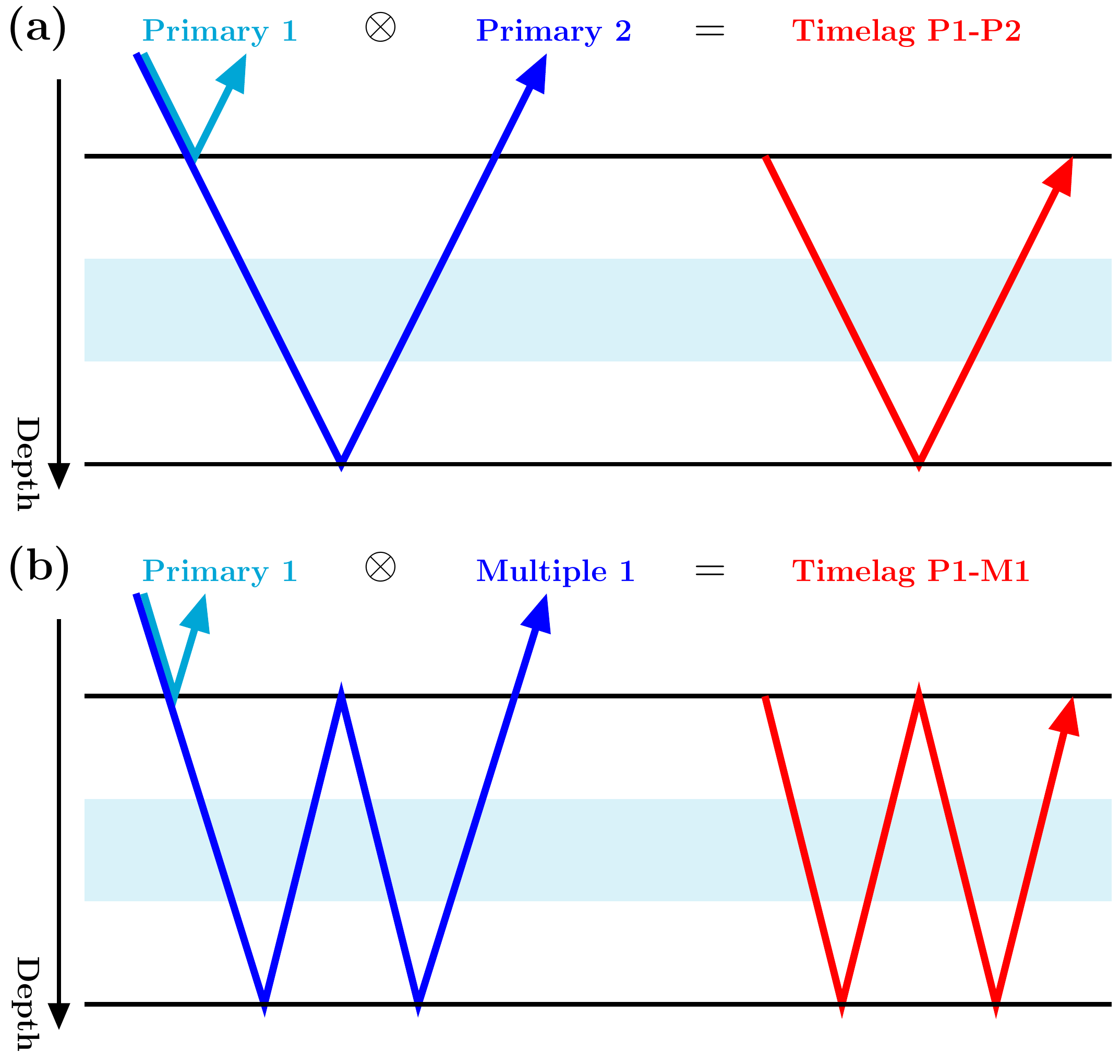}
\caption{Example of cross-correlations of primary 1 with primary 2 (a) and multiple 1 (b). In the resulting time-lags, the common path is canceled, i.e. the overburden effects are effectively removed. Note, how the multiple travels through the reservoir layer (in light blue) an additional time compared to primary 2.}
\label{fig:CrossCorrelations}
\end{figure}

Once an isolated reflection response of the primary and multiples around the reservoir is acquired, the traveltime differences in the reservoir between the baseline and monitor survey can be determined. However, the time-lapse delays due to changes in the overburden have not yet been accounted for. To remove these delays a primary reflection ($P1$) above the reservoir is identified, which can then be used as a control event that includes the overburden time differences, but excludes the differences in the reservoir. Identifying such an event is trivial in the isolated response, which is solely comprised of the target zone reflections. Next, a second reflection, either a second primary ($P2$) or an internal multiple ($M1$, $M2$ etc.), is identified, this time from below the reservoir so that it contains both overburden and the reservoir traveltime differences. Subsequently, these two reflections are cross-correlated to remove overburden changes \citep{vanIJsseldijk2023}:
\begin{equation}
\label{eqn:correlation}
C_{\star}(\mathbf{x}_0,\tau) =
\int_0^\infty \Theta_{P1}(t+\tau) R_{b}(\mathbf{x}_0,t+\tau) \Theta_{\star}(t) R_{b}(\mathbf{x}_0,t) d t.
\end{equation}
Here, $C$ represents the cross-correlation, $\mathbf{x}_0$ the zero-offset coordinate, and $\Theta$ is a time-window that selects the desired reflection from the isolated reflection response as follows:
\begin{equation}
\label{eqn:theta}
\Theta_{\star}(t) = \begin{cases}
1, \text{if } t_{\star}-\epsilon < t \leq t_{\star} + \epsilon \\
0, \text{otherwise.}
\end{cases}
\end{equation}
The subscript $\star$ denotes either primary 1 or 2 ($P1$ or $P2$) or an internal multiple, $t_\star$ then specifies the twt of this event, and $\epsilon$ serves as a small shift to include the full wavelet of the reflection data. Figure \ref{fig:CrossCorrelations}a and b show how Figure \ref{eqn:correlation} is used to retrieve the time-lags of primary 1 with primary 2 as well as with multiple 1, respectively. The final results after cross-correlation no longer contain time-lapse overburden effects. The next step, where the baseline and monitor time-difference will be computed, will, therefore, only contain time-differences from the reservoir and none from the overburden. \\
Finally the actual time-lapse differences are retrieved. In order to achieve this, the time-lag correlations for both the baseline and monitor study are computed using Figure \ref{eqn:correlation}. Thereafter, these time-lags are cross-correlated once more to retrieve the traveltime difference ($\Delta t_{\star}$) in the reservoir:
\begin{equation}
\label{eqn:extraction}
\Delta t_{\star}(\mathbf{x}_0) = \argmax_\tau \left( \int_0^\infty C_{\star}(\mathbf{x}_0,t+\tau) \bar{C}_{\star}(\mathbf{x}_0,t) d t \right).
\end{equation}
$C_\star$ is the retrieved correlation from Figure \ref{eqn:correlation}, the bar denotes retrieval in the monitor survey as opposed to the baseline survey. The argument of the maximum is used to find the zero-offset time shifts from the final correlation. \Citet{vanIJsseldijk2023} apply this method to a synthetic  set to find time differences in three subsurface dome structures. Application of the Marchenko method to field data is more complex due to strict amplitude requirements on the reflection response \citep[][]{Staring2018,Brackenhoff2019}. The next section discusses how to to overcome this and other limitations on field data. 

\par\null

\section*{Application to the Troll field}
\begin{figure}[tb!]
\includegraphics[width=.5\textwidth]{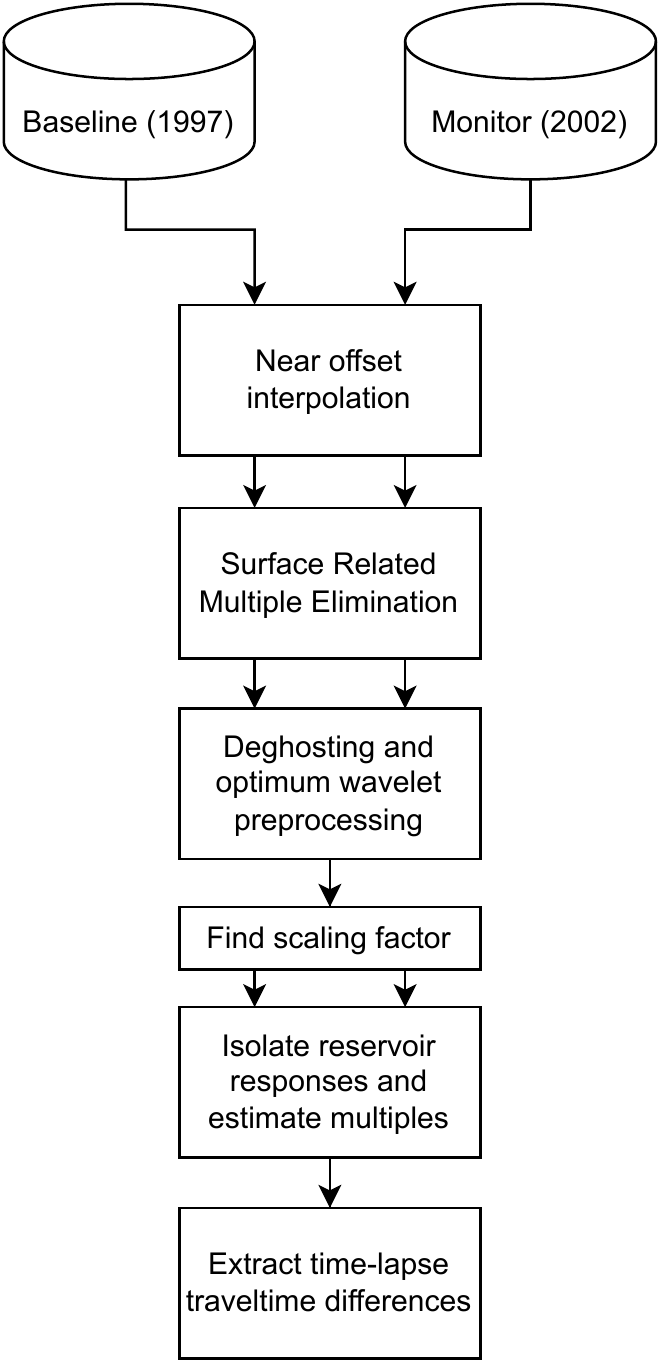}
\caption{Flowchart detailing the different processing steps for retrieving traveltime differences of the Troll time-lapse data. }
\label{fig:Flow}
\end{figure}

The methodology is applied to a marine time-lapse data set shot over the Troll Field off the Norwegian coast. In 1997 a 3D baseline survey was conducted over the Troll West Gas Province, followed by a monitor survey in 2002. This study considers a 2D subset of these 3D surveys. The time-lapse target is a hydrocarbon-water contact. Specifically the hydrocarbon is a gas layer underlain by an oil leg with a varying thickness between 0 and 28 m \citep{Hellem1986}. The contact partially coincides with a geologic structure, which makes extracting time-lapse effects from the data challenging \citep{Bannister2005}. Additionally, the repeatability of the surveys is subpar, further complicating time-lapse analysis with conventional methods \citep{QuVerschuur2020}. \\
A number of basic pre-processing steps were applied to both datasets. First, it should be noted that the data were not completely raw, namely some unknown time gain and wavelet processing as well as far offset muting was performed.
The known pre-processing first applied a regularisation to get a 2D geometry with 481 co-located source and receiver positions sampled at 12.5 m. Near-offsets of about 85 m were interpolated by parabolic Radon transform \citep{Kabir1995}. Next, surface-related multiple elimination (SRME) was applied to get a clearer image of the reservoir reflections \citep{Verschuur1992}. Note that this does not handle internal multiples, which will be dealt with separately using the Marchenko-based isolation. Deghosting was then applied as well as optimum wavelet processing to ensure zero-phase character. \\
Aside from the unknown scaling factors, application of the Marchenko method also lacked a velocity model and suffered from the limited recording time of 2 s. To properly remove internal multiples with the Marchenko method, it is important that the scaling of the reflection response is accurate \citep{vanderNeut2015b}. Moreover, a (smooth) velocity model is required to compute an estimate of the twt between the focal depth and the surface. The limited recording time means that some of the internal multiples from the target zone are not recorded. In the following sections these three problems will be taken care of one-by-one. Finally, the results of isolating the reservoir response and extracting time-differences will be discussed. An overview of the different processing steps is given in \autoref{fig:Flow}.

\subsection*{Velocity model estimation}
\begin{figure}[tb!]
\includegraphics[width=\textwidth]{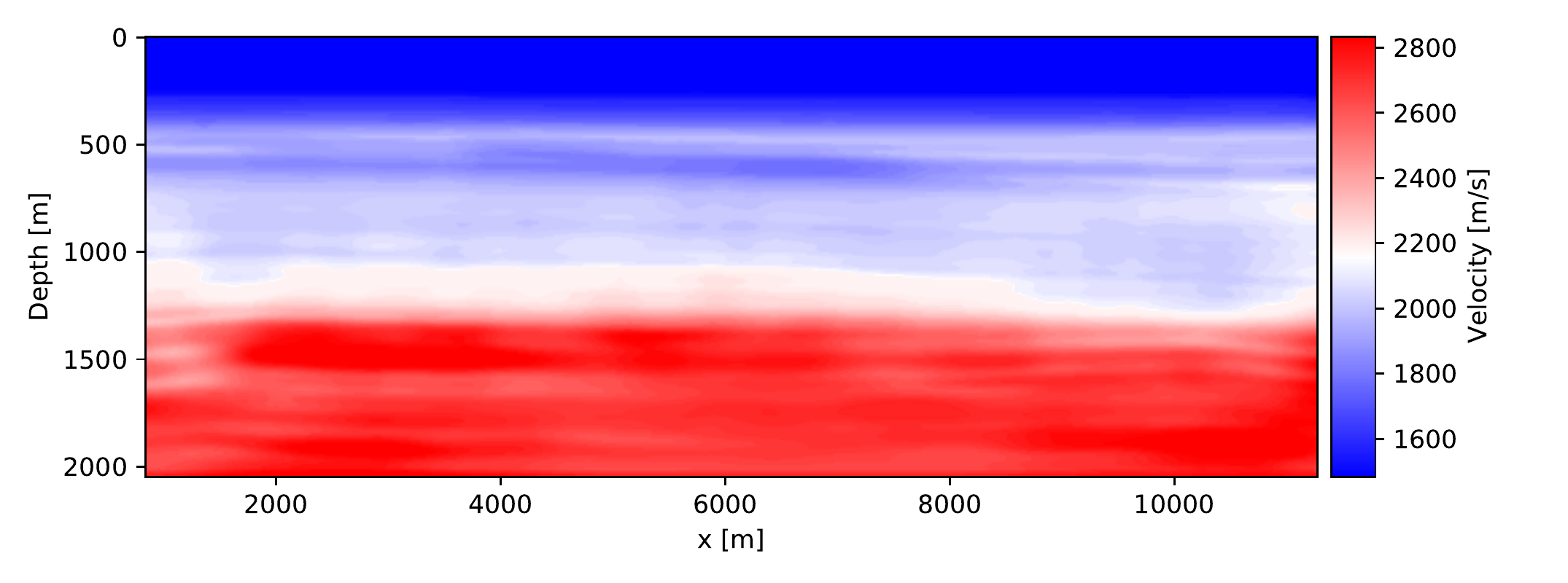}
\caption{Baseline velocity model of the Troll field, derived from \citet{QuVerschuur2020}. }
\label{fig:velmod}
\end{figure}

In order to separate focusing functions from Green's functions an estimation of the twt from the surface to the focal depth is required. In general, this is achieved with the use of an eikonal solver in a smooth velocity model. 
\citet{QuVerschuur2020} use a simultaneous joint migration inversion (S-JMI) approach to find the approximate baseline velocity model. However, this model was not readily available for this study. Instead the model was derived from Figure 14a in \citet{QuVerschuur2020}. By matching the RGB values in the figure with the colour bar, a rough estimate of the original  model was acquired, as shown in Figure \ref{fig:velmod}. Since no major velocity changes are expected between the two time-lapse surveys, this model can be used for the Marchenko-based isolation of both the reservoir in the baseline and the monitor survey. 

\subsection*{Scaling of the reflection response}
\citet{Brackenhoff2016} introduces a cost function that can be minimised to find the optimal scaling of the reflection data. Here, this function has to be slightly adapted in order to handle the extrapolated functions, but the principle of the method remains the same. The principle of using these cost functions relies on the fact that the energy in the redatumed reflection response usually decreases due to the removal of internal multiples. Consequently, if the scaling of the data is too low not all multiple energy will be removed. Whereas, if the scaling is too high the energy will be excessively magnified. Hence, only if the data are correctly scaled the cost function will be minimized due to the lower energy by the multiple removal. The redatuming in Equation \ref{eqn:redatum} is relatively expensive to apply multiple times on the data for each scaling factor. Therefore, the computationally inexpensive alternative of double dereverberation (ddr) by double-focusing is considered instead \citep{Wapenaar2021}:
\begin{equation}
\label{eqn:doublefocus}
R_{ddr}(\xs{R}',\xs{S}',t) =  \int_{\mathbb{S}_0} v^+(\xs{R},\xs{R}',t) * U^{-,+}(\xs{R},\xs{S}',t)   d \xs{R}.
\end{equation}
The downside of Equation \ref{eqn:doublefocus} is that some remaining interactions of the overburden will still be present in the computed reflection response $R_{ddr}$. As mentioned before the upside is that the double-focusing method is relatively cheap, more stable and can easily be applied for a wide range of scaling factors \citep{Staring2018}. The cost function considers the ratio of the energy in the reflection response before and after Marchenko redatuming, and is applied as follows: 
\begin{equation}
\label{eqn:costfunction}
J (b) = \frac{\left|\left|R_{ddr}(\xs{R}',\xs{S}',t) \right|\right|_2}{\left|\left|R_{ddr,0}(\xs{R}',\xs{S}',t) \right|\right|_2}.
\end{equation}
Here, $||...||_2$ denotes the L2-norm, $J$ is the cost function, and $b$ is the scaling of the original reflection response. $R_{ddr,0}$ is the response for the focusing and Green's function of the first Marchenko iteration, which parallels a standard time-reversal experiment without internal multiple removal \citep{Wapenaar2017partII}.  $R_{ddr}$ is the response after the final Marchenko iteration, as specified by the user. \\ 
The estimation of the scaling factor depends on the removal of internal multiples by the Marchenko method. This process is complicated by the limited recording time of 2 s. This means that instead of picking a deep focal level below all reflectors, as is ideal \citep{Brackenhoff2016}, a shallower focal level has to be used to calculate the cost functions. Consequently, the cost functions do not contain a lowest minimum, and only a range of possible factors is acquired. This range is further refined to find a single scaling factor by inverting for the reflection response in Equation \ref{eqn:redatum} for a limited amount of scaling factors, and visually examining the resulting reflection response. Finally, a single factor of $5\cdot 10^{-5}$ is found using this method for both time-lapse surveys.

\subsection*{Multiple retrieval beyond recorded time}
\begin{figure}[tb!]
\includegraphics[width=\textwidth]{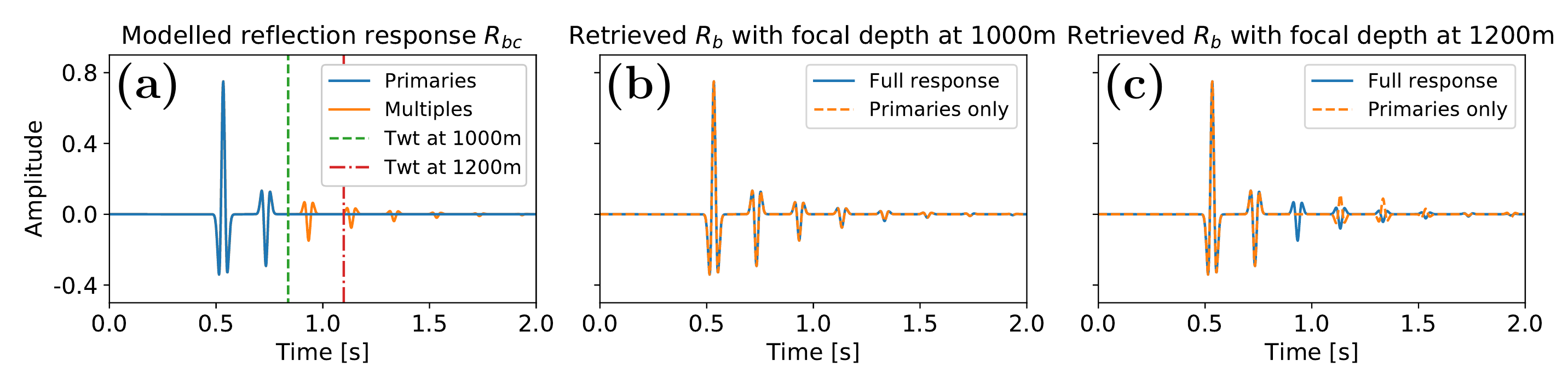}
\caption{Reflection response $R_{bc}$ modelled in a medium with a velocity and density profile of 1500 [m/s] \& [kg/m$^3$] for depths 0 to 400 m, $4000$ [m/s] \& [kg/m$^3$] from 400 to 800 m and 1750 [m/s] \& [kg/m$^3$] below 800 m. In (a), the primary reflections are shown in blue and the internal multiples in orange, with the two-way traveltime (twt) for depths 1000 m and 1200 m marked with a green dashed and red dot-dashed line, respectively. Next, the retrieved (by MDD) reflection response above the focal depth $R_b$ are shown for Marchenko with the full response vs for the primary reflections only in blue and orange, respectively. Note that, in (b) and (c), response $R_b$ can be properly retrieved (using Equation \ref{eqn:isolation}) from data that contains all events within the time-window (b), if any events are missing in this window response $R_b$ can no longer be correctly retrieved (c).}
\label{fig:MagicMults}
\end{figure}

The short recording time not only constrains the effectiveness of the scaling factor, but also causes that some of the target multiples are not recorded in the data. These multiples provide complementary information of the target zone and are ideally recovered from the data. A closer look at the focusing functions in Equation \ref{eqn:isolation} reveals that these functions are solely defined between $t=0$ and the two-way traveltime to the focal depth (i.e. in the truncated medium). Because of this finite behaviour, the focusing functions do not require all the recorded internal multiples to retrieve the reflection response without underburden. Consequently, Equation \ref{eqn:isolation} can be used to compute the internal multiples outside of the recorded time. 
Note that this only applies for underburden removal (i.e. Equation \ref{eqn:isolation}), but not for overburden removal (i.e. Equation \ref{eqn:redatum}) as the Green's functions are infinite in time, hence all events in the Green's functions are used to reconstruct the reflection response without overburden. \\
In order to illustrate how the focusing functions can be used to retrieve additional multiples outside the recorded time, a simple 3-layer 1D model is considered. The acoustic impedance contrasts in this model are very strong, ensuring that a strong multiple train is generated as shown in Figure \ref{fig:MagicMults}a. Next, two focal levels are considered; one at 1000 m that does not include any multiples in it's response, and one at 1200 m that does. Subsequently, the Marchenko method and Equation \ref{eqn:isolation} are used to retrieve a reflection response free of underburden effects, once with the full reflection response, and once using solely the primary reflections (i.e. with limited information). On the one hand, shown in Figure \ref{fig:MagicMults}b in the case where no multiples are present in the time-window at the focal depth, the underburden is correctly removed independent on inclusion of all multiples in the original reflection response. On the other hand if the time window does include one or more multiples, the underburden can no longer accurately be removed when using primary reflections only (e.g. Figure \ref{fig:MagicMults}c). This is caused by the fact that using the primaries only in Figure \ref{fig:MagicMults}c, reconstructs the reflection response with incomplete data (i.e. it misses a multiple important for reconstruction). It is, therefore, important that the time-window contains all information of both primaries and multiples when removing the underburden. \\ 
This numerical experiment suggests that even though the internal multiples may not be recorded, they can still be extracted from the data by using the focusing functions. However, one has to ensure that all information is included in the time-window, meaning that focal depth and corresponding time-window should be picked as closely to the end of the recorded data as possible, in order to ensure all events for finding the focusing functions are included. Since the focal level in Equation \ref{eqn:isolation} can be arbitrary chosen at any level below the two primary reflectors, this constraint can easily be satisfied in the current study.

\subsection*{Results of the Marchenko-based isolation}
\begin{figure}[tb!]
\includegraphics[width=\textwidth]{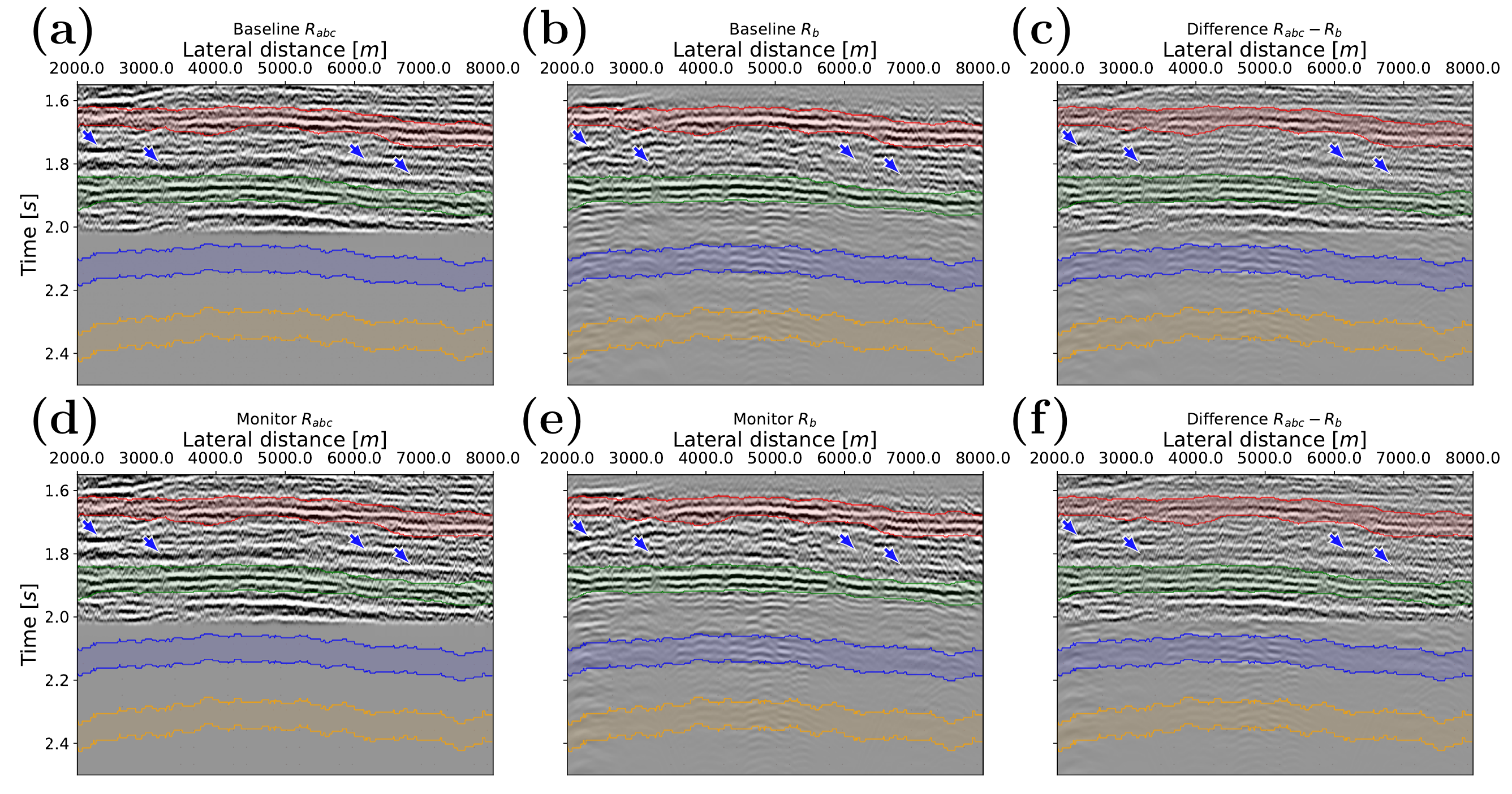}
\caption{Zero-offset gathers of the baseline survey before (a), after Marchenko-based isolation (b) and their difference (c). The second row shows the same gathers for the monitor survey. The red, green, blue and orange highlights mark the first and second primary as well as the first and second order multiples from these primaries. The arrows point at removed multiples originating from the overburden. As shown in (a) and (d) the original data only recorded 2 s, but the Marchenko method is able to retrieve multiples beyond this cut-off (i.e. b and e). }
\label{fig:Z0panels}
\end{figure}

\begin{figure}[tb!]
\includegraphics[width=\textwidth]{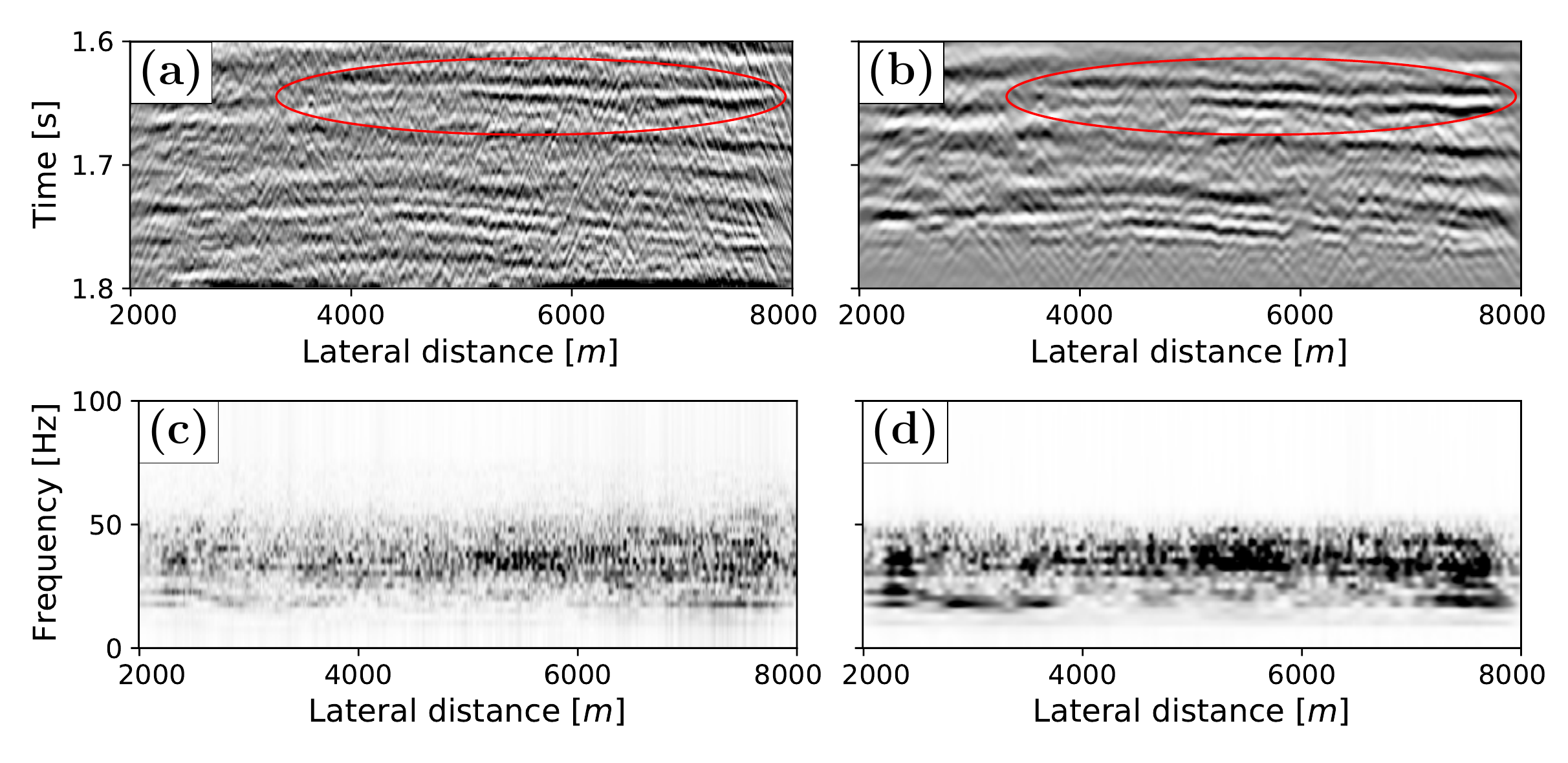}
\caption{Time-lapse differences between the stacked baseline and monitor images, zoomed in on the target zone. The difference before Marchenko-based isolation is shown in (a). After Marchenko-based isolation (b) the reflectivity differences are a lot clearer as marked by the red ellipse. The frequency spectra for panel (a) and (b) are shown in (c) and (d), respectively. }
\label{fig:DiffStacks}
\end{figure}

After the data are properly pre-processed as described in the previous sections, the Marchenko-based isolation can now be applied. The result of this isolation is shown in Figure \ref{fig:Z0panels}. A reflector right above the reservoir is selected as primary 1 just after $1.65$ s twt. The first focal depth, for overburden removal, is chosen above this reflector at a depth of $1575$ m. Next, primary 2 is identified around $1.9$ s twt, with a third primary following closely behind. Hence, the second focal depth, for underburden removal, is picked right in between these two reflectors at a depth of $1975$ m. Subsequently, primary 1 and primary 2 can be isolated from the full reflection response. Finally, the internal multiples of the target zone are enhanced by increasing the amplitudes of the coda of the down-going focusing function before MDD of Equation \ref{eqn:isolation} as described in \citet{vanIJsseldijk2023}. \\
Figure \ref{fig:Z0panels} shows the results for the baseline and monitor study in the first and second row, respectively. Note that this figure shows zero-offsets gathers with red, green, blue and orange highlights for the first primary, second primary, first multiple and second multiple of the target area, respectively. These windows will later on be used to extract time-differences. The first column in the figure shows the original reflection response with a recording time until $2$ s. There are no data available at $2.15$ s and $2.4$ s, where the internal multiples are expected. In the second column the response after over- and underburden removal is shown. Note that the internal multiples can now be observed at times beyond $2$ s twt, these multiples are especially strong at a lateral distance of $3500$ m to $6000$ m. Furthermore, the over- and underburden reflections are removed not only below and above the focal levels, but also inside the target zone as marked by the blue arrows in the figure. Finally, the third column shows the difference between the first two panels, once again the removal of overburden multiples in the target area between primary 1 and 2 is noted. However, there also seems to be quite some coherent information removed from the first and second primary. This could be an indication that the optimal scaling factor has not been found or another explanation could be that the MDD applied a correction on the phase of the signal \citep[e.g.][]{vanDalen2015}. \\
The results in Figure \ref{fig:Z0panels} for the baseline and monitor survey are quite similar, although some minor differences can be detected when carefully analyzing the panels on each row. In order to more precisely compare the two studies, a raw stack of all shots in the reflection data was computed for both the regular and the isolated response. The results of these stacks and their spectra are displayed in Figure \ref{fig:DiffStacks}. The time-lapse effects are especially strong in the first primary at $1.65$ s twt. Additionally, the stack created from the isolated response in (b) is much cleaner compared to the stack of the full reflection response in (a). Consequently, the isolated stack shows more continuity, which will aid in a better interpretation of the data. 

\begin{figure}[tb!]
\includegraphics[width=\textwidth]{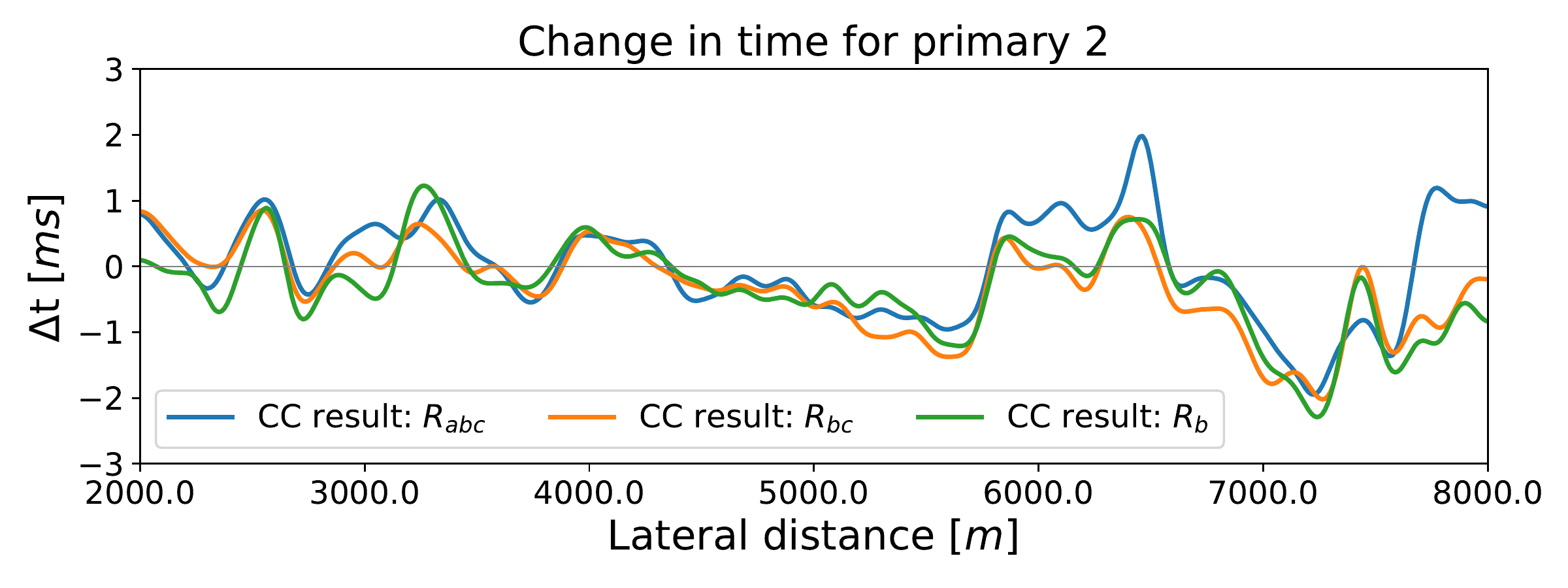}
\caption{Time difference between the baseline and monitor time-lag of primary 1 and primary 2. The blue, orange and green colours represent the time-shift in the zero-offset response before Marchenko, after overburden removal and after full isolation, respectively. }
\label{fig:P1Deltas}
\end{figure}

\subsection*{Extracting time-lapse traveltime differences}
\begin{figure}[tb!]
\includegraphics[width=\textwidth]{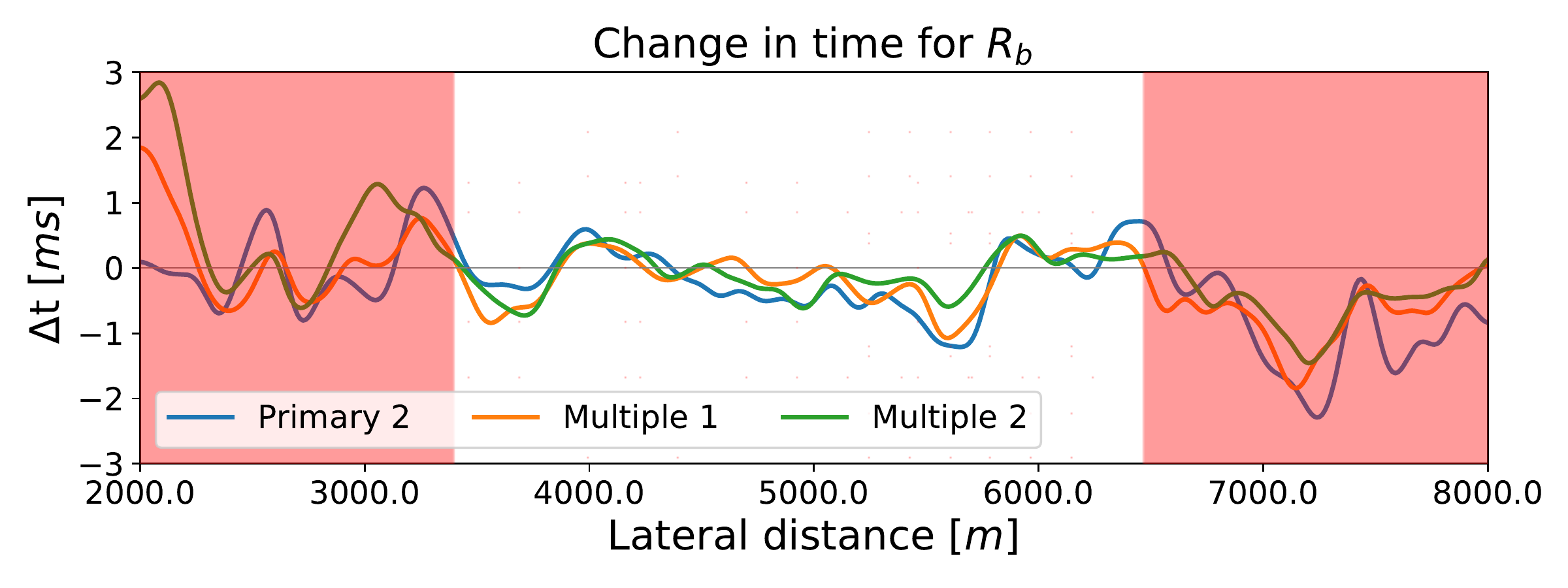}
\caption{Time difference between the baseline and monitor time-lag. The time-lags between primary 1 and, respectively, primary 2 (blue), multiple 1 (orange) and multiple 2 (green) are shown. Note that the time shifts have been divided by 2 and 3, respectively, for multiple 1 and 2 to compare them with the primary shift. These differences where calculated in the isolated response, the red shading indicates the areas where the multiples are weaker, as shown in Figure \ref{fig:Z0panels}b.}
\label{fig:RbDeltas}
\end{figure}

In order to accurately retrieve the time-lapse traveltime differences, the reflection the time is interpolated in the frequency domain from 4ms to 1ms. Next, the coloured windows in Figure \ref{fig:Z0panels} are now used to extract the time-lapse differences as described in Equation \ref{eqn:extraction}. Primary 1, highlighted in red, is correlated either with the second primary or with one of the multiples giving the time-lag between the two events. The baseline timelag is then correlated with the monitor timelag to find the time difference. First, the timelag of $P1$ and $P2$ is considered for the full reflection response $R_{abc}$, the response after overburden removal $R_{bc}$ and the response after total isolation of the target zone $R_b$. The results of this experiment are shown in Figure \ref{fig:P1Deltas}. Note that the three lines are matching quite closely aside from the edges, where the deviations are slightly larger. The close match is easily explained with Figure \ref{fig:Z0panels}, which shows that both primaries are already visible in the original reflection response with little obstructions from the overburden. The worse performance at the edges is most likely due to edge effects introduced by the MDD. \\
Figure \ref{fig:RbDeltas} shows the retrieved traveltime differences for the primary and multiples in the isolated response $R_b$, which is the only response that contains the predicted internal multiples for this analysis. The red shading indicates the zone where the multiples are weak in amplitude shown in Figure \ref{fig:Z0panels} (b) and (e). The results for multiple 1 (in orange) and multiple 2 (in green) have been divided by a factor of 2 and 3, respectively, in order to have a fair comparison with the results from primary 2 (in blue), as these multiples probe the reservoir two or three times. Especially, between the red zones a strong match between the results of primaries and multiples in Figure \ref{fig:P1Deltas} and the results in Figure \ref{fig:RbDeltas} is noticed, which implies that the multiples are successfully estimated outside the recording time. The Marchenko-based isolation in Equation \ref{eqn:isolation} predicts multiples based on only primary reflection data. This also happens in target area $b$, the primaries are used to predict the internal multiples between $P1$ and $P2$. \\
The extracted time-differences are compared to the results by \citet{QuVerschuur2020}, who retrieved approximate velocity changes. Some similarities are observed when looking at the sign of the change in velocity in Figure 15 of \citet{QuVerschuur2020} and the differences in Figures \ref{fig:P1Deltas} and \ref{fig:RbDeltas}, that is the negative time-shifts coincide with an increase in velocity observed by \citet{QuVerschuur2020}. Additionally, based on approximations of the thickness of the reservoir (around 30 m) and the velocity changes (between -20 and 40 m/s), the time-lapse traveltime differences are roughly estimated to lie somewhere between $+1$ ms and $-1.5$ ms, which is similar to the result found in this study. However, it is difficult to make one-on-one comparisons between the two studies as they consider different quantities that are not directly related to each other. 

\par\null

\section*{Discussion}

As shown in the previous section, Marchenko-based isolation of the reservoir response demonstrates promising results to improve time-lapse monitoring on field data. The reflection data are better defined (i.e. overburden interactions are reduced), hence providing a superior image of the target zone, and time-lapse traveltime differences can be extracted from the primaries and estimated multiples. In this section, some limitations and potential future advancements of the method will be discussed. \\
Firstly, the current method only considers one time-lapse aspect of the data, namely traveltime differences. Traditional time-lapse studies, oftentimes, also consider changes in the amplitudes in the form of amplitude versus offset or angle (AVO and AVA) analyses. Recent studies investigated how angle-dependent reflectivity can be obtained with the Marchenko method for lateral invariant media with constant velocities \citep{Alfaraj2020}. If this method can be further extended to include fully heterogeneous media, it will be able to provide new insights for Marchenko-based time-lapse monitoring with AVA. \\ 
These AVA time-lapse analyses generally consider changes in both P- and S-waves \citep{Landro2001}. On the contrary, the Marchenko method is mostly used for the acoustic case only, because the causality constraint is no longer ensured when elastic waves are introduced \citep{daCosta2014,Reinicke2020}. This further complicates advanced AVA analysis on an isolated response, where only P-waves are considered. \\
The biggest obstacle to apply the Marchenko method to field data is the strict scaling requirements on the reflection data.   This either calls for a carefully designed preprocessing scheme \citep{DukalskiReinicke2022}, which is not always feasible in the case of legacy data. Alternatively, a cost function can be minimised to find the optimal scaling; the approach that was considered in the current study. This technique suffered from the limited recording time of the data, causing most events to fall outside this limit. The cost function, therefore, was unable to converge to a minimum value. \\
Furthermore, the limited recording time also meant that the internal multiples of the target zone were not recorded. It was shown that these multiples could, in theory, be estimated accurately with the Marchenko method. The accuracy of this prediction on field data is harder to determine, where missing multiples of the target zone were predicted based on the primaries. While these multiples were not recorded in the data itself, once they were estimated with the Marchenko method, they still provided valuable insights into the time lapse differences of the target zone. \\
The current study extracted time-lapse differences from the zero-offset gathers. Instead, multiple offsets can be combined to acquire traveltime differences. On the one hand, the advantage of using multiple offsets is that the result is smoother over multiple offsets and thus more robust. On the other hand, the disadvantage is that the correlation windows have to be manually picked for each individual shot in the reflection response. While the inclusion of multiple offsets was explored for this study, it was found that the results did not significantly improve, hence the zero-offset gathers were deemed sufficient for the time-lapse analysis. However, computation of velocity changes from the traveltime differences demands multiple offsets to be used.

\par\null

\section*{Conclusion}

The purpose of this study was to apply Marchenko-based isolation of the reflection response on marine time-lapse data of the Troll field, in order to extract traveltime differences between the baseline and monitor surveys. To achieve this goal an approximate velocity model was acquired, and a suitable scaling of the reflection response was determined. Subsequently, the over- and underburden were removed by twice employing Marchenko redatuming, once above and once below the target zone. From this newfound response two primaries and multiples were identified, and the time-lag between these events calculated. Finally, the baseline and monitor time-lags were correlated to obtain the time-lapse traveltime differences in the reservoir. \\
The methodology successfully eliminated imprints from signal originating outside of the target zone, resulting in an unobstructed view of the reservoir reflections. However, the target zone was relatively free of multiple reflections to begin with. It would, therefore, be interesting to test the methodology on a data with stronger interference from internal multiples (such as a subsalt reservoir), to be able to conclusively determine the impact of the time-lapse Marchenko scheme. On the current data the method was able to restore internal multiples lying outside of the recorded times. These were used together with the unobstructed primaries to retrieve the time-lapse traveltime differences inside of the reservoir. \\
These results open the door for future time-lapse applications of the Marchenko method, which ultimately can aid in our understanding of time-lapse changes in a reservoir caused by storage or production of resources inside the subsurface. 

\par\null

\section*{Acknowledgements}

The authors thank Equinor (formerly Statoil A.S.) for providing the marine time-lapse datasets of the Troll Field. We are also grateful for insightful discussions with Eric Verschuur and Christian Reinicke. This research was funded by the European Research Council (ERC) under the European Union’s Horizon 2020 research and innovation program (grant agreement no. 742703).

\par\null

\section*{Data availability}

The algorithms associated with this research are available and can be accessed via the following URL: \href{https://gitlab.com/geophysicsdelft/OpenSource}{https://gitlab.com/geophysicsdelft/OpenSource}. The Troll field data are currently not publicly accessible, but are available upon reasonable request and with the permission of Equinor. The synthetic 1D example in \autoref{fig:MagicMults} is available at: \href{https://github.com/Ohnoj/Geophysics/tree/main/Marchenko1D}{https://github.com/Ohnoj/Geophysics/tree/main/Marchenko1D}. 

\par\null

\section*{Conflict of interest}

The authors declare no conflict of interest.

\par\null

\selectlanguage{english}
\FloatBarrier
\bibliography{bibliography/TL_bib.bib}


\newpage
\listoffigures
\processdelayedfloats  


\end{document}